# Superconducting PdTe Thin Film Via Topotactic Transformation, Toward Topological Superconductors

Hee Taek Yi[1,*], Min Ge[2], Renjie Xie[3], Colby J. Stoddard[1], David H. Yi[4], Xiaoyu Yuan[1], Xiong Yao[3], and Seongshik Oh[1,*]

[1] *Department of Physics & Astronomy, Rutgers University, Piscataway, New Jersey 08854, USA*
[2] *The Instruments Center for Physical Science, University of Science and Technology of China, Hefei 230026, China*
[3] *Ningbo Institute of Materials Technology and Engineering, Chinese Academy of Sciences, Ningbo 315201, China*
[4] *Department of Physics, Mellon College of Science, Carnegie Mellon University, Pittsburgh, Pennsylvania 15213, USA*

*E-mail: taeggy@physics.rutgers.edu and ohsean@physics.rutgers.edu

**Topological superconductors (TSCs) hosting Majorana zero modes (MZMs) offer a pathway to fault-tolerant quantum computation. PdTe is a promising TSC candidate due to its topological surface states and a reasonable superconducting critical temperature of ~4.5 K. However, it has been challenging to grow PdTe thin films with bulk-like superconducting properties. Here, we show that high-quality, superconducting PdTe thin films can be grown using molecular beam epitaxy (MBE). The films exhibit a sharp superconducting transition ($T_{onset}$ = 4.43 K with transition width of 0.06 K), comparable to that of bulk crystals. This was made possible via a topotactic transformation from a $PdTe_2$ buffer layer to a PdTe phase by growing Pd on top under Te-deficient conditions. Structural and transport analyses confirm the NiAs-type structure of PdTe, as well as its two-dimensional superconducting behavior and excellent air stability. These findings suggest that the MBE-grown PdTe films and their heterostructures are a promising platform for topological superconductivity and Majorana physics.**

Topological superconductors (TSC), which host Majorana zero modes (MZM), have emerged as a promising platform for fault-tolerant computation relying on their non-Abelian statistics.[1-3] TSCs can be realized by combining a conventional superconductor with a topological surface state, such as topological insulators, Weyl/Dirac semimetals with Fermi arcs, or magnetic materials. There have been three pathways toward achieving TSC states: a) search for intrinsic topological superconductors such as Fe(Te,Se), b) engineering the proximity effect that combines superconductors (SC) with topological surface states in a heterostructure form, and c) utilizing the magnetic proximity effect in an interface between an s-wave SC and magnetic materials to create triplet superconductivity.[4-12]

PdTe has recently emerged as a promising new candidate for an intrinsic TSC. It possesses two essential ingredients for TSCs: superconductivity and topological surface states. The transition temperature of PdTe at 4.5 K is significantly higher than that of $PdTe_2$. Furthermore, PdTe exhibits Dirac semimetal features with a bulk Dirac cone and topological surface states with Fermi arcs. The absence of strongly correlated 3d elements also suggests that the pairing mechanism is likely to be conventional electron-phonon coupling rather than exotic mechanisms such as spin fluctuations [13-17] Angle-resolved photoemission spectroscopy (ARPES) measurements on PdTe single crystals have revealed the coexistence of a fully gapped superconducting state on the surface and a gapless nodal state in the bulk. Specific heat measurements support this coexistence. Meanwhile, thermal conductivity measurements point toward multiple nodeless gaps[13-16]. PdTe also offers significant advantages for TSC via the proximity effect in heterostructures. Its hexagonal crystal structure, with an in-plane lattice constant of 0.415 nm, closely matches those of a topological insulator, $Bi_2Se_3$, and an altermagnet, MnTe.[18-21] Accordingly, high-quality heterostructures with a clean interface between PdTe and these related materials can be envisioned toward proximity-induced TSCs. All these properties make PdTe an attractive candidate for realizing TSCs, either intrinsically or in heterostructures. However, it has been challenging to grow high-quality superconducting PdTe thin films,[22-24] with bulk-like $T_c$ reported only recently on vacuum-annealed PLD (Pulsed Laser Deposition)-grown films.[25]

In this study, we report the synthesis of high-quality PdTe thin films, using MBE (Molecular Beam Epitaxy) technique. Transport measurements exhibit a high onset temperature ($T_{onset}$ = 4.43 K) with $T_0$ = 4.37 K and a residual resistivity ratio (*RRR*) of 10, consistent with the reported $T_c$ of bulk crystals.[14-16] We successfully achieved the PdTe phase by depositing Pd in Te-deficient conditions on a $PdTe_2$ buffer layer via a topotactic transformation, a structural phase change mediated by atomic rearrangement.[26-29] Comprehensive characterizations, including in situ Reflection High Energy Electron Diffraction (RHEED), X-ray Diffraction (XRD), low-temperature electric transport measurements, and Scanning Transmission Electron Microscopy (STEM), confirm the high crystallinity with $T_c$ continuously tunable between those of $PdTe_2$ and PdTe. These films also allowed the first Scanning Tunneling Microscopy (STM) studies on the PdTe system, answering key unresolved questions in this system, which will be reported in a follow-up paper.

We grew PdTe thin films in a custom-built MBE system with a base pressure of $\sim 10^{-10}$ Torr, on $10\times10\times0.5$ $mm^3$ $Al_2O_3$ (0001) substrates, which provide good lattice matching with PdTe, when in-plane rotation is taken into account (Figure *S*1, Supporting Information). The RHEED images in Figure 1a, monitored during the film growth, present the 6-fold in-plane symmetry and the alternating spacings of $\sqrt{3}d'$ and $d'$ in two high-symmetry directions. The sharp streaks indicate the high structural quality of the film. The $\sqrt{3}d'$ RHEED spacing in PdTe aligns well with the in-plane spacing, $2d$, of $Al_2O_3$. Rutherford backscattering spectroscopy (RBS) shows a 1:1 ratio for Pd:Te, confirming the stoichiometry of the grown film (See Figure *S*2, Supporting information).

Figure 1b exhibits the temperature-dependent longitudinal sheet resistance, $R_{xx}$, of the PdTe film on a semi-log plot. A sharp superconducting transition occurs at $T_{onset}$ = 4.43 K with zero resistance temperature, $T_0$ = 4.37 K, accompanied by a residual resistivity ratio, *RRR*, of 10, consistent with the reported $T_c$ of bulk crystals.[14-17] This sharp transition, $T_{onset}$ - $T_0$ = 0.06 K, combined with a high *RRR* of 10, indicates the high-quality epitaxial growth of the PdTe film. The inset displays the strong air stability of $T_c$. We remeasured the PdTe thin film kept in air for 3 months (red symbols and line) and found that the degradation is minimal, $T_{c,\ as\text{-}grown}$ - $T_{c,\ aged} < 0.1$ K. This air stability is a strong advantage of PdTe compared to other non-oxide superconductors such as Fe(Te,Se) for both research and applications. For instance, $T_c$ of Fe-based superconductors can easily degrade even in a few days without a proper capping layer.[30] We observed a large anisotropy of $B_{c2}$ between out-of-plane and in-plane directions, as highlighted in Figure 1c. $B_{c2}$ along the in-plane direction is 10 times larger than that along the out-of-plane direction. The high anisotropy, compared to the bulk value of 1.3, demonstrates that the thin film PdTe behaves like a 2D superconductor due to dimensional confinement.[13] The $B_{c2}$ data along the two different directions were fitted using the multiband Werthamer-Helfand-Hohenberg (WHH) model.[13, 31]

This approach was necessary because the single-band WHH model, the standard approach for conventional superconductors, fails to fit $B_{c2}$ in both directions simultaneously. This yielded a diffusivity ratio of $\eta$ = D2/D1 = 0.21, where the $D_1$ and $D_2$ denote the diffusivities of each band. This indicates a strong asymmetry in quasiparticle transport between the two bands, consistent with the bulk case (See Figure *S*3, Supporting Information).[13] Band structure calculations and magneto-transport studies on PdTe single crystals previously provided signatures of non-trivial band structure topology and the presence of two-electron and two-hole pockets in this system.[17] The Hall effect measurements taken at 7 K on our PdTe thin films are also consistent with this previous study taken on the single crystals. he extracted mobilities are $\mu_{h1}$ = 434 cm$^2$/V·s, $\mu_{h2}$ = 2,885 cm$^2$/V·s, $\mu_{e1}$ = 343 cm$^2$/V·s, and $\mu_{e2}$ = 2,573 cm$^2$/V·s. Corresponding carrier densities are $n_{h1}$ = 7.3 × 10$^{20}$/cm$^3$, $n_{h2}$ = 2.6 × 10$^{19}$/cm$^3$, $n_{e1}$ = 9.8 × 10$^{20}$/cm$^3$, and $n_{e2}$ = 1.8 × 10$^{19}$/cm$^3$. (see Figure *S*4, supporting information).

Previously, thin film growth techniques such as MBE and sputtering led to only the $PdTe_2$ phase. $PdTe_2$ belongs to the $CdI_2$-type crystal structure ($P\bar{3}m_1$), and PdTe belongs to NiAs-type ($P6_3/mmc$). $PdTe_2$ and PdTe share an identical in-plane crystal structure with slightly different lattice constants of 4.07 Å and 4.15 Å, respectively.[16, 32] One notable difference between them is that $PdTe_2$ has a van der Waals bonding along the c-direction, whereas PdTe has a covalent/ionic bonding between Pd and Te. Moreover, $PdTe_2$, being the end-member in the Pd-telluride phase diagram, can be readily grown with an adsorption-controlled self-limited growth mode. On the other hand, being surrounded by other competing phases, such as $Pd_3Te_2$ and $Pd_2Te_3$, it does not seem possible to grow PdTe in such a self-limited growth mode.

Not surprisingly, in our initial trials of direct growth of PdTe on $Al_2O_3$ substrate, we observed 3D spots with faint streaks in the RHEED patterns (not shown here), indicating poor crystallinity and mixed phases of the grown films. On the other hand, we were able to grow high-quality $PdTe_2$ films under a Te-rich growth condition with minimal effort, so we decided to use $PdTe_2$ as a buffer layer to grow PdTe on top. After preparing the highly crystallized $PdTe_2$ thin film on the $Al_2O_3$ substrate, we deposited Pd and Te on the $PdTe_2$ buffer layer with a nominal Te/Pd flux ratio of one. The RHEED spacing, *d*, showed an intermediate value between those of PdTe and $PdTe_2$. The transport measurement showed a $T_c$ of 2.7 K, which is between the bulk $T_c$ of 4.5 K for PdTe and ~1.8 K for $PdTe_2$.[14, 16, 17] This observation suggests that it is likely to be an intermediate or a mixed phase between PdTe and $PdTe_2$.

To understand the growth mechanism and achieve high-quality PdTe films, we performed a systematic study on various Te/Pd ratios of *r,* as shown in Figure 2. After preparing a 32 nm-thick $PdTe_2$ buffer layer, we deposited Pd and Te with the same amount of Pd used for the buffer layer but with different *r* values ranging from 3 to 0. Here, the *r*0 sample implies Pd growth without any Te supplied. Figure 2a shows that the RHEED spacings on the two high symmetry directions match well with those of $PdTe_2$. With decreasing *r*, *d* shortens, indicating broadening of the in-plane lattice constant. With the *r* value reaching 0.25 (i.e. Te/Pd = 0.25), the lattice constant saturates at 4.14 Å. Interestingly, a third-order surface reconstruction peaks begin to emerge for $r \leq 0.25$, which will be discussed later.

The XRD patterns in Figure 2b also exhibit phase transformation from $PdTe_2$ to PdTe with decreasing *r*. All the XRD peaks in the *r*3 sample can be identified as $PdTe_2$ (00n) and $Al_2O_3$ (00n) peaks. As *r* decreases, PdTe (00n) peaks appear and coexist with the $PdTe_2$ (00n) peaks. When *r* reaches 0.3, the $PdTe_2$ (00n) peaks vanish entirely, and only PdTe (00n) peaks remain. However, upon closer inspection of the XRD data (Figure 2c), the PdTe (004) peak appears at a slightly lower angle than the known bulk value, as indicated by the dashed line in Figure 2c. This could be due to the presence of some interstitial Te species, and as the Te flux is further reduced to $r = 0.25$, this peak eventually shifts to the bulk value.

The normalized longitudinal resistance as a function of temperature, shown in Figure 3a, reveals how the superconducting transition temperature evolves as the film transforms from $PdTe_2$ to PdTe. Except for the *r*3 sample, which is shown above to be $PdTe_2$ ($T_c \approx \sim 1.8$ K), all the other samples exhibit $T_c$ above 1.9 K. Figure 3b summarizes the data for all the samples. Bulk-like $T_c$ (4.4 K) of the PdTe phase is observed for $r \leq 0.25$, but as *r* increases from 0.25 to 0.3, $T_c$ abruptly drops to 2.7 K, and *RRR* also drops sharply, reaching a minimum at *r*0.5. With a further increase in *r*, *RRR* rebounds and reaches 15 at the *r*3 sample (See Figure *S*5, Supporting Information, for further details). The sudden drop in *RRR* between *r*0.25 and *r*0.3 is likely due to the development of atomic-scale disorder, such as the interstitial Te in the PdTe phase, as suggested above based on the XRD data in Figure 2.

These observations strongly suggest that by growing Pd under Te-poor growth conditions on top of $PdTe_2$, the entire film, including the $PdTe_2$ buffer layer, transforms into the PdTe phase. This implies that Pd can readily diffuse through the film at the growth temperature, as visualized in the schematic of Figure 3c. Under low *r* conditions, excess Pd atoms diffuse into the $PdTe_2$ buffer layer and bond with neighboring Te atoms within the van der Waals gap, converting $PdTe_2$ into PdTe. Such a process is generally referred to as a topotactic transformation. Even if growth begins with the $PdTe_2$ buffer, high Pd diffusivity facilitates the topotactic transformation to convert $PdTe_2$ into a high-quality PdTe film for $r \leq 0.25$. Interestingly, the surface reconstruction peaks observed in the RHEED of *r*0.25 samples serve as a barometer indicating that the film has been completely transformed into PdTe. Notably, the *r*0 sample exhibits a slightly lower $T_c$ and *RRR* compared to the *r*0.25 sample. This could be due to the presence of Pd clusters on the surface of the *r*0 sample, as confirmed by Atomic Force Microscope (AFM) images in Figure *S*6 (Supporting Information). In Fig. 3, it is notable that while $T_c$ and *RRR* correlate with each other for $r < 0.5$, such is not the case for $r > 0.5$. This is because on the $r > 0.5$ side, even though the reduction in disorder (i.e., increase in RRR) helps enhance the superconductivity, the $T_c$ should eventually drop toward that (1.8 K) of $PdTe_2$.

To further investigate how the topotactic transformation occurs, we performed STEM studies on three representative samples: *r*3, *r*0.5, and *r*0.25. As shown in Figure 4, the STEM images of the *r*0.25 and *r*3 samples display the NiAs-type (PdTe) and $CdI_2$-type ($PdTe_2$) stacking patterns, respectively. A distinct van der Waals gap is clearly visible in the *r*3 sample. The measured lattice constants along the c-axis are 5.7 Å for *r*0.25 and 5.2 Å for *r*3, consistent with the known values of PdTe and $PdTe_2$.[15, 17, 32] In contrast to the homogeneous structures observed in the *r*3 and *r*0.25 samples, the *r*0.5 sample displays mixed stacking patterns in Figure 4b, which reveals a clear boundary between the two phases, PdTe (top) and $PdTe_2$ (bottom). The fact that we observe a sharp superconducting transition in all these films at temperatures between that of PdTe (4.4 K) and $PdTe_2$ (1.8 K) suggests that the PdTe/$PdTe_2$ heterostructures form transparent interfaces with strong proximity coupling. However, a closer inspection of STEM images near the interface between the PdTe and $PdTe_2$ phases reveals that the detailed structures in each region deviate from those of the pure PdTe and $PdTe_2$ phases. The stacking angle varies from ~47° to ~50° in PdTe and from ~54° to ~52° in $PdTe_2$ near the interface, corresponding to local out-of-plane strains of ~4.6% tensile for PdTe and ~2.6% compressive for $PdTe_2$, highly localized at the PdTe/$PdTe_2$ interface.

## CONCLUSIONS

In conclusion, we have developed the first high-quality, superconducting PdTe thin films. This was made possible by depositing Pd under Te-poor growth conditions on a PdTe2 buffer layer via a topotactic transformation. The optimized films exhibited a bulk-like superconducting transition with a narrow transition width of 0.06 K and a high *RRR* of ~10. Comprehensive structural analyses, including XRD, RBS, AFM, and RHEED, revealed a clear phase evolution between $PdTe_2$ and PdTe as a function of the Te/Pd ratio. Interestingly, over the entire flux range, only $PdTe_2$, PdTe, and their mixtures - without any other second phases such as $Pd_2Te_3$ or $Pd_3Te_2$ – were observed in the films. With high-quality PdTe films available, it is now feasible to probe previously inaccessible topological properties, such as Majorana Zero modes, in this novel topological superconductor candidate. Moreover, the 2D superconducting behavior and exceptional air stability make this system a promising platform for topological or magnetic superconductor heterostructures, thereby facilitating the realization of proximity-induced TSCs.

## MATERIALS AND METHODS

**Thin-film growth.** We prepared PdTe films on 10×10×0.5 $mm^3$ $Al_2O_3$ (0001) using a custom-built MBE system (SVTA) with a base pressure of $\sim 10^{-10}$ Torr. $Al_2O_3$ substrates were cleaned ex-situ by UV-generated ozone and in situ by heating to 750 °C under oxygen pressure of $5 \times 10^{-7}$ Torr. Before depositing the PdTe, we employed a $PdTe_2$ buffer layer. $PdTe_2$ buffer layer and PdTe thin film were grown at 275 °C and 250 °C, respectively. High-purity Pd and Te were thermally evaporated using Knudsen diffusion cells for film growth. All source fluxes were calibrated in situ using a quartz crystal microbalance.

**Transport measurements.** All transport measurements were performed using the van der Pauw geometry, except for the upper-critical-field measurements. For upper-critical-field measurements, a Hall-bar pattern was prepared by optical lithography, with a channel length of 25 μm and a width of 30 μm. Keithley 2400 source-measure units and 7001 switch matrix system were controlled by a LabView program for longitudinal and Hall resistance measurements. A Physical Property Measurement System (PPMS, Quantum Design Inc.) was employed to cool down to 2 K.

**HAADF-STEM and XRD measurements.** High-angle annular dark-field scanning transmission electron microscopy (HAADF-STEM) characterizations were performed at the Instruments Center for Physical Science in China. TEM samples were prepared by a focused ion beam (Helios G5, Thermo Fisher Scientific) using 2.0 keV $Ga^+$ ions for final milling. HAADF STEM images were acquired with 300 keV electrons with collection angles ranging from 67 to 275 mrad using a double aberration-corrected JEOL.

**Competing interests**

Authors declare that they have no competing interests.

**Data availability**

All data are available in the manuscript or the supplementary information and are available from the corresponding author upon request.

**Author Contributions**

S.O. and H.T.Y. conceived the experiments. H.T.Y. performed the thin-film growth and transport measurements. R.X. and X.Yao. performed XRD. M.G. performed STEM. D.H.Y. performed AFM. C.S. and X.Yuan. analyzed and fitted the transport data. S.O. and H.T.Y. analyzed the data and wrote the paper with feedback from other co-authors.

**Funding Sources**

This work is supported by the National Science Foundation's DMR 2451900, the National Natural Science Foundation of China (Grant Nos. 12304541), and the Hundred Talents Program of Chinese Academy of Sciences.

**Acknowledgements**

We would like to thank Jewook Park and Hoyeon Jeon at Oak Ridge National Laboratory for their helpful discussions and STM work, which will be reported in a follow-up paper.

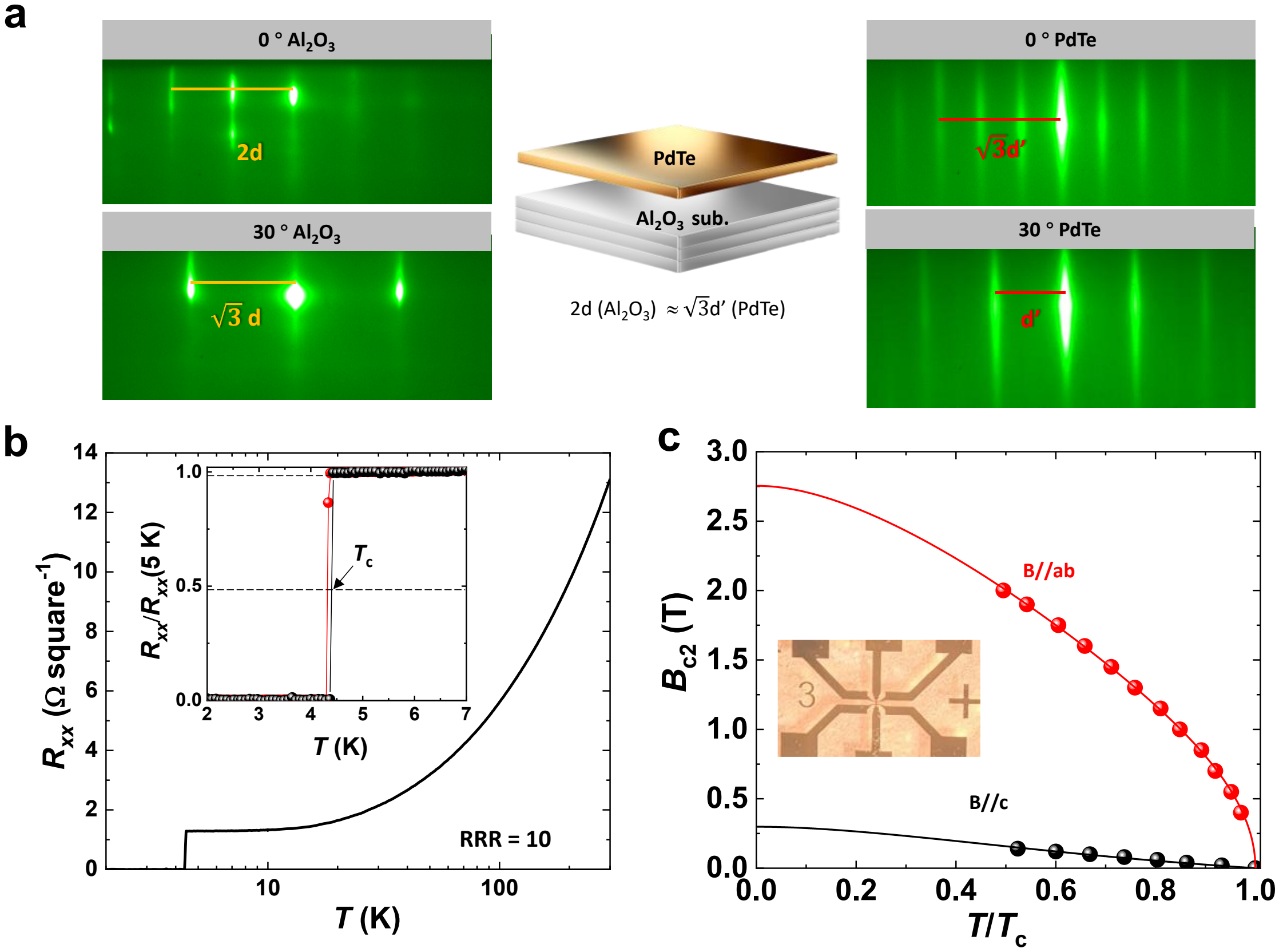

**Figure 1.** RHEED and transport properties of a PdTe film. (a) RHEED patterns along two high-symmetry directions for a 40 nm-thick PdTe film (right panels) and an $Al_2O_3$ substrate (left panels). A lattice match is observed between $2d$ and $\sqrt{3}d$' (see Figure *S*1, Supporting Information). (b) semi-log plot of the temperature-dependent longitudinal sheet resistance, $R_{xx}$, of a PdTe film. The inset shows an enlarged graph of the normalized $R_{xx}$ versus temperature, measured on a 10 × 10 mm$^2$ sample using the van der Pauw geometry. The dashed line indicates half the normal state resistance. Red symbols and line represent the remeasured $R_{xx}$(T) after 3 months of air exposure. (c) Upper critical fields measured along in- and out-of-plane directions. Symbols represent experiment data: black for out-of-plane and red for in-plane. Solid lines show fitted lines using the two-band WHH model. The inset displays the Hall bar-patterned sample used for the upper critical fields measurements. The channel area measures 25 μm in length and 30 μm in width.

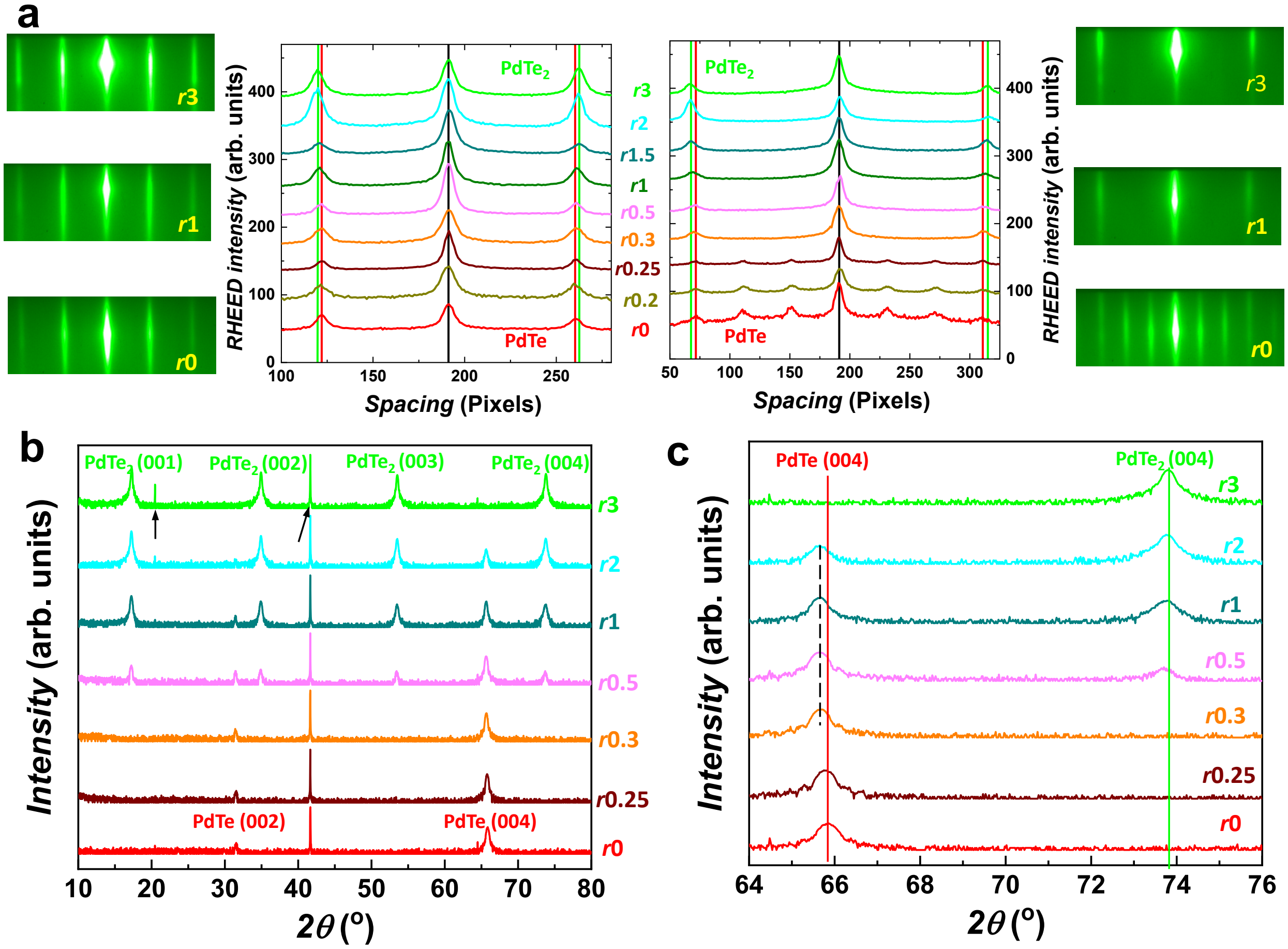

**Figure 2.** RHEED spaces and XRD analysis of Pd-telluride samples with varying *r* values. (a) As the *r*-value decreases, the RHEED spacing reduces and saturates at *r* = 0.25. At this point, 1/3-order surface reconstruction peaks begin to emerge. The surface reconstruction peaks work as a barometer, indicating the completion of the topotactic transformation. (b) evolution of the (*00n*) XRD peaks as a function of *r*-values. The black arrows indicate the (*00n*) $Al_2O_3$ substrate peaks. (c) enlarged view of the (*004*) peaks corresponding to PdTe and $PdTe_2$. Red and green vertical lines indicate the PdTe and $PdTe_2$ (*004*) peak angles. The dashed black line shifted to a smaller angle suggests an enlarged *c*-axis lattice constant of PdTe in samples with intermediate *r* values.

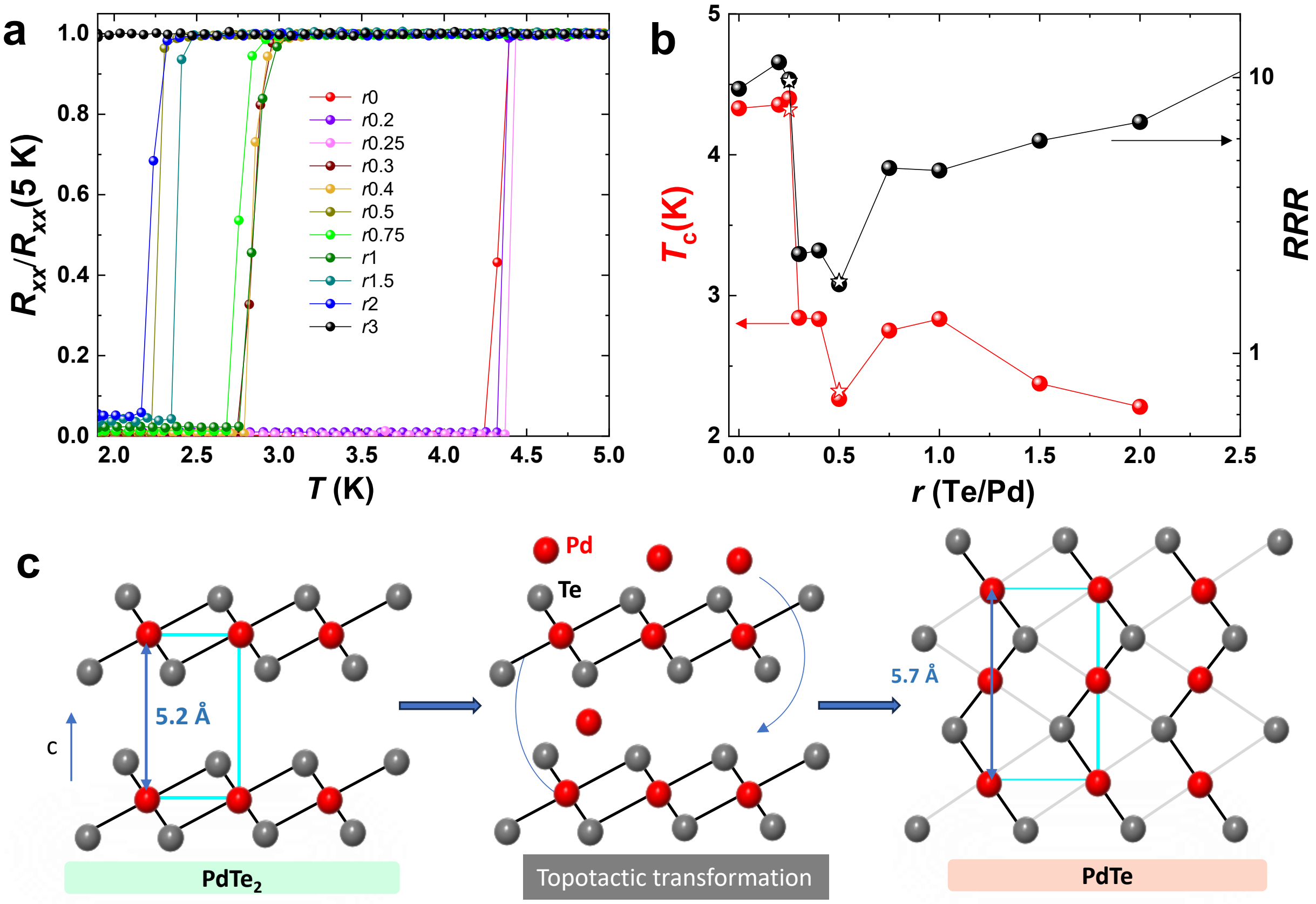

**Figure 3.** Transport properties of Pd-telluride films with varying $r$ values and a schematic of the topotactic transformation. (a) Normalized $R_{xx}$ as a function of temperature for the samples as $r$ varies. (b) Summary of $T_c$ and $RRR$ for the measured samples. Star symbols represent another sample prepared under an identical condition, demonstrating the reproducibility of the synthesis method. (c) A schematic of the topotactic transformation from $PdTe_2$ to PdTe, driven by the diffusion of excess Pd ions in samples with $r < 1$.

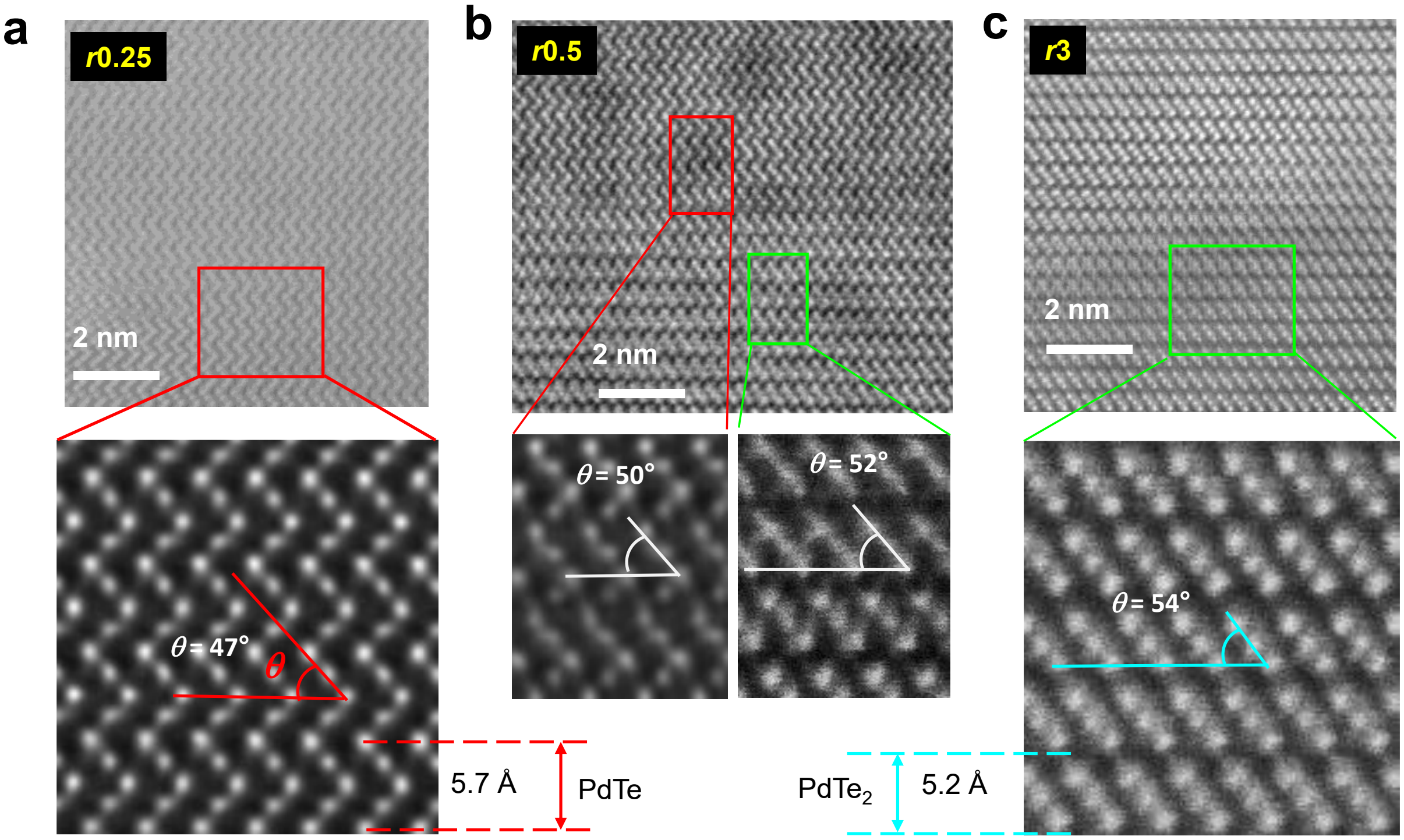

**Figure 4.** STEM images of Pd-telluride films with varying $r$ values. (a) $r$ = 0.25 sample shows the PdTe structure with a stacking angle of 47°. (b) $r$ = 0.5 sample exhibits two distinct regions: the upper region displays a PdTe-like stacking with an angle of 50°, while the lower region shows a $PdTe_2$-like stacking with an angle of 52°, presumably due to some compositional mixing between the two phases. (c) $r$ = 3 sample presents a stacking angle of 54°, characteristic of $PdTe_2$.

Supporting information

# Superconducting PdTe Thin Film Via Topotactic Transformation, Toward Topological Superconductors

Hee Taek Yi[1, *], Min Ge[2], Renjie Xie[3], Colby J. Stoddard[1], David H. Yi[4], Xiaoyu Yuan[1], Xiong Yao[3], and Seongshik Oh[1, *]

[1] *Department of Physics & Astronomy, Rutgers University, Piscataway, New Jersey 08854, USA*
[2] *The Instruments Center for Physical Science, University of Science and Technology of China, Hefei 230026, China*
[3] *Ningbo Institute of Materials Technology and Engineering, Chinese Academy of Sciences, Ningbo 315201, China*
[4] *Department of Physics, Mellon College of Science, Carnegie Mellon University, Pittsburgh, Pennsylvania 15213, USA*

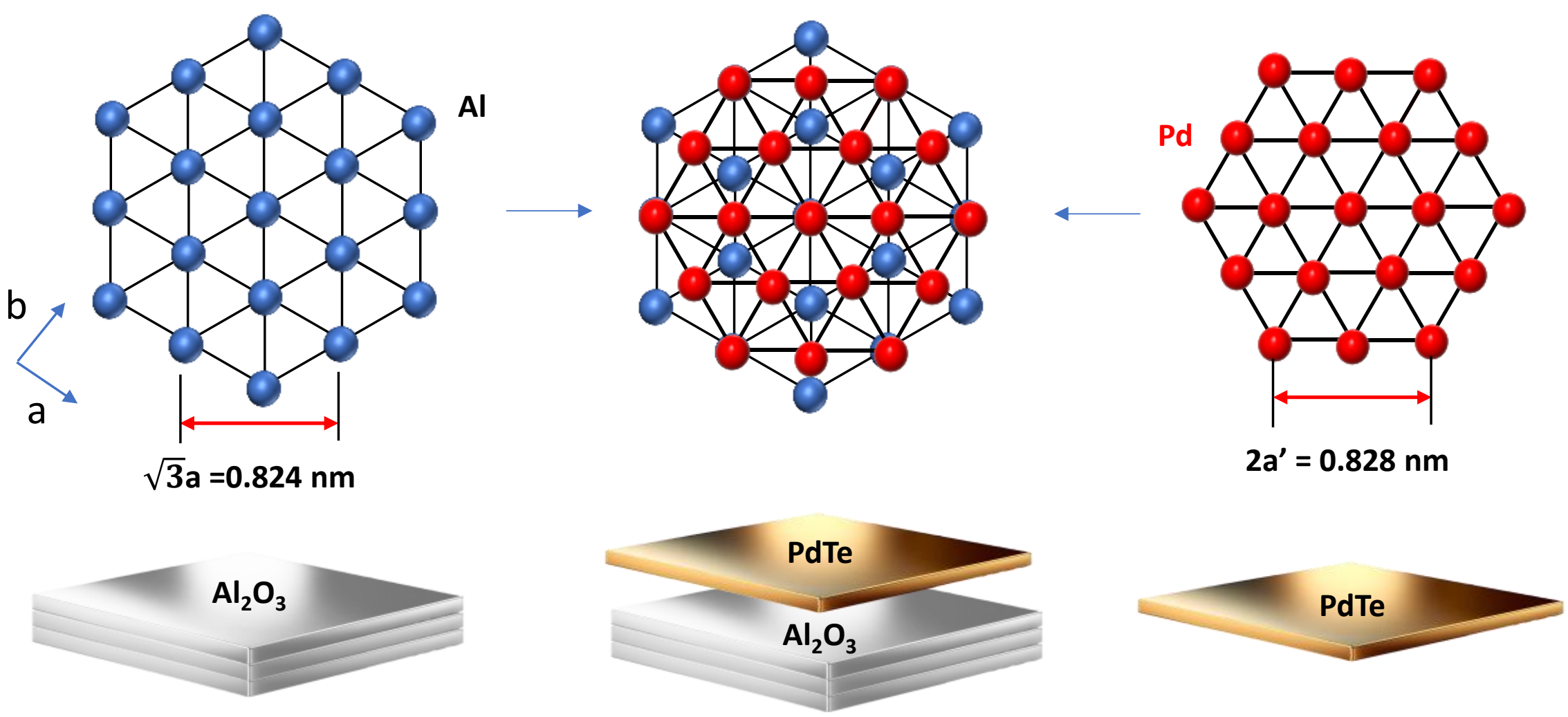

**Figure *S*1**. Schematic illustration of rotational commensurate lattice matching. Blue and red symbols represent Al and Pd atoms, respectively. The illustration highlights the commensurate relationship between the in-plane lattice constant of $Al_2O_3$ and that of PdTe.

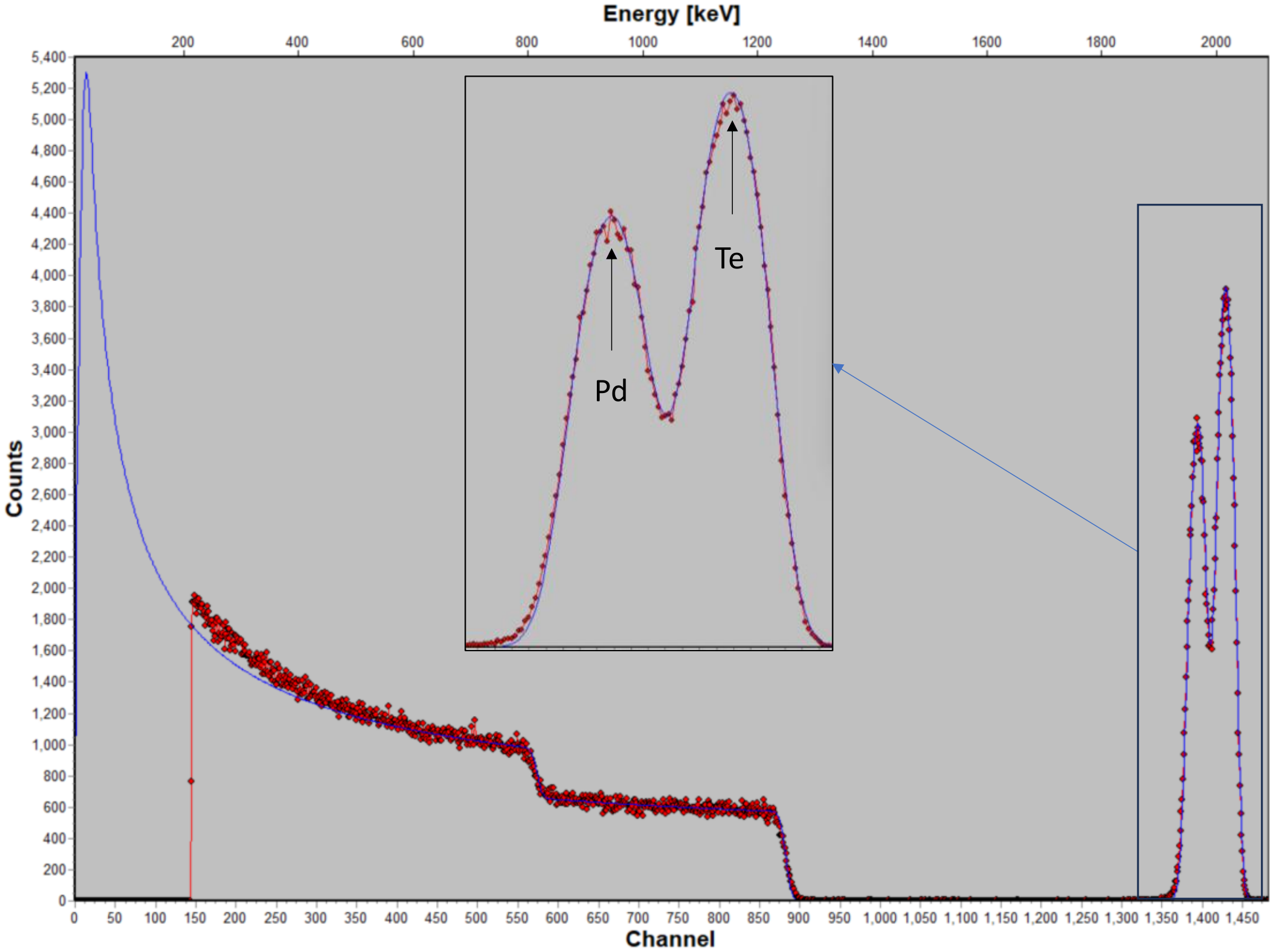

**Figure *S*2**. Rutherford backscattering spectroscopy (RBS) spectrum recorded on a PdTe thin film. RBS experimental data for the *r*0.25 PdTe film (red circles) are compared with simulated data (blue line). The inset shows an enlarged view of the spectrum, highlighting the Pd and Te signals. The simulation corresponds to a stoichiometric composition of PdTe with a 1:1 ratio.

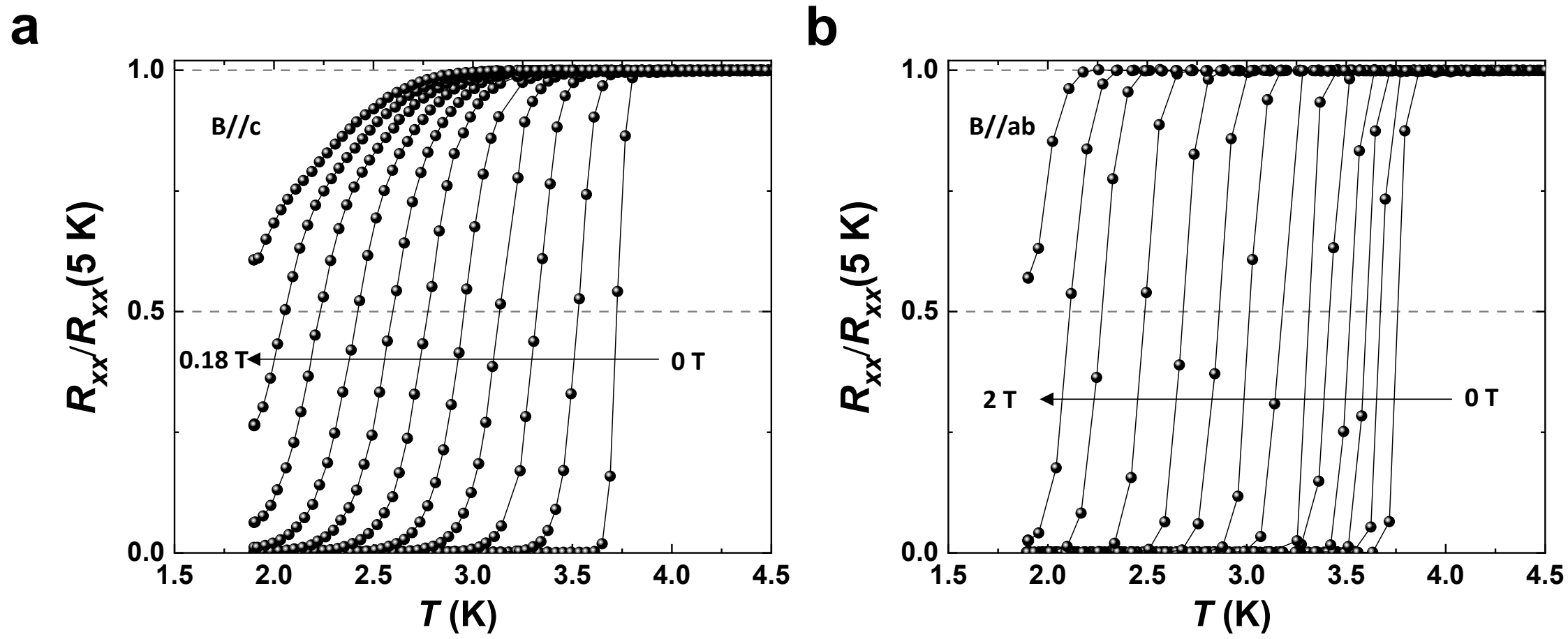

**Figure *S*3.** Temperature-dependent $R_{xx}$ of the PdTe thin film under varying magnetic fields along (a) perpendicular and (b) parallel direction to the ab plane.

The upper critical field is fit to the two-band Werthamer-Helfand-Hohenberg (WHH) model defined by

$$\frac{w}{\lambda_0}\big(\ln t + U(h)\big)\big(\ln t + U(\eta H)\big) + (1 + \lambda_-/\lambda_0)\big(\ln t + U(h)\big) + (1 - \lambda_-/\lambda_0)\big(\ln t + U(\eta h)\big) \quad [Eq. 1]$$

where $\mathrm{U}(x) = \psi\left(x + \frac{1}{2}\right) - \psi\left(\frac{1}{2}\right)$, $\mathrm{t} = \mathrm{T}/T_c$, $\eta = D_2/D_1$, $\mathrm{w} = \lambda_{11}\lambda_{22} - \lambda_{12}\lambda_{21}$, $\lambda_- = \lambda_{11} - \lambda_{22}$, $\lambda_0 = \sqrt{\lambda_-^2 - 4\lambda_{12}\lambda_{21}}$, and $\psi(x)$ is the digamma function.[1] The parameters $D_1$ and $D_2$ are the intraband diffusivities of bands 1 and 2, respectively. The constants $\lambda_{11}$ and $\lambda_{22}$ quantify the intraband superconducting coupling, while $\lambda_{12}$ and $\lambda_{21}$ quantify the interband coupling. For thin films, [2,3]

$$h = \frac{D_1}{4\pi kT}\left(2eH_{c,2}\,sin\,\theta + \frac{1}{3\hbar}\left(deH_{c,2}\,cos\,\theta\right)^2\right) \quad [Eq. 2]$$

where $H_{c,2}$ is the upper critical field, $-$e is the electron charge, k is Boltzmann's constant, d is the effective superconducting film thickness and θ is the angle from the film plane. We used the coupling constants for single-crystal PdTe found in Ref.[4] and found the best fit for η, $D_1$, and d. Note that $\lambda_{12}$ and $\lambda_{21}$ only appear in Eq. 1 as the product $\lambda_{12}\lambda_{21}$, so we only considered the product $\lambda_{12}\lambda_{21}$. Table S1 shows the fit parameters for a simultaneous fit of the in-plane and out-of-plane upper critical fields.

| $\sqrt{\lambda_{12}\lambda_{21}}$ | $\lambda_{22}/\lambda_{11}$ | η | $D_1$ $(m^2/s)$ | d (nm) |
|---|---|---|---|---|
| 0.627 | 1.32 | 0.21 | $3.66 \times 10^{-3}$ | 12.5 |

**Table *S*1.** Fit parameters to Eq. 1 for simultaneous fit to in-plane and out-of-plane upper critical fields.

The longitudinal and transverse conductance converted from the measured longitudinal and transverse resistances with the channel ratio, *l*/*w* =1.2 are defined as follows:

$$\sigma_{xx} = \frac{\rho_{xx}}{\rho_{xx}^2+\rho_{xy}^2} \text{ and } \sigma_{xy} = \frac{-\rho_{xy}}{\rho_{xx}^2+\rho_{xy}^2} \text{ with } \rho_{xx} = \frac{R_{xx}}{1.2} \text{ and } \rho_{xy} = R_{xy}$$

The conductivity of an N-carrier system is given by[5]

$$\sigma_{xx} = \sum_{i=1}^{N} \frac{n_i q_i \mu_i}{1+\mu_i^2 B^2} \text{ and } \sigma_{xy} = \sum_{i=1}^{N} \frac{s_i n_i q_i \mu_i^2 B}{1+\mu_i^2 B^2}$$

where $B$ is the applied magnetic field and $n_i, q_i$, and $\mu_i$ are the carrier density, charge, and mobility of the *i*th carrier, respectively. The coefficient $s_i$ is $-1$ for electrons and $+1$ for holes. We performed a simultaneous least-squares fit to $\sigma_{xx}$ and $\sigma_{xy}$ to determine the density, mobility, and type of each carrier. The two- and three-carrier models yielded poor fits to the experimental data, whereas the four-carrier model provided an excellent fit as shown in Fig. *S*4. The extracted mobilities are $\mu_{h1}$ = 434 $cm^2$/V·s, $\mu_{h2}$ = 2,885 $cm^2$/V·s, $\mu_{e1}$ = 343 $cm^2$/V·s, and $\mu_{e2}$ = 2,573 $cm^2$/V·s. Corresponding carrier densities are $n_{h1}$ = 7.3 × $10^{20}$/$cm^3$, $n_{h2}$ = 2.6 × $10^{19}$/$cm^3$, $n_{e1}$ = 9.8 × $10^{20}$/$cm^3$, and $n_{e2}$ = 1.8 × $10^{19}$/$cm^3$.

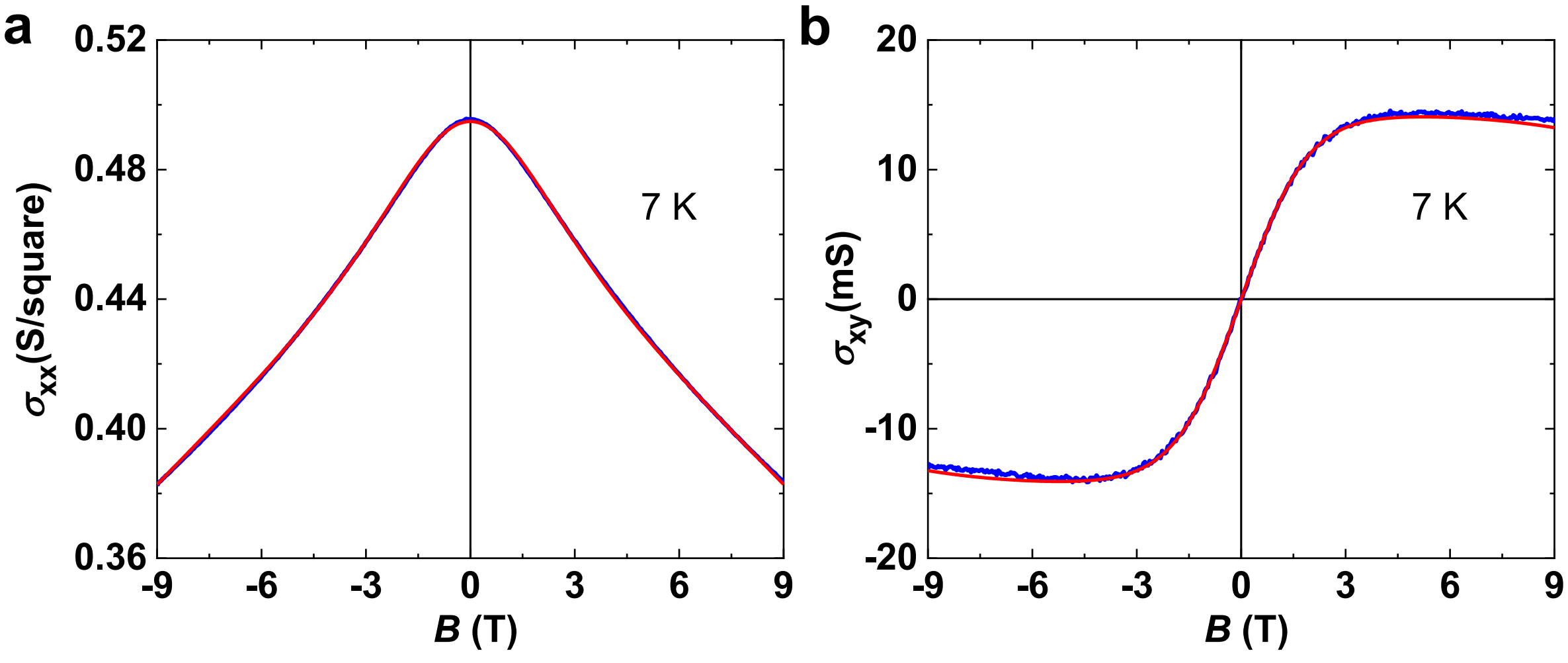

**Figure *S*4.** Hall effect measurement at 7 K in the normal state (a) longitudinal and (b) Hall conductance as a function of applied magnetic fields. Black and red curves represent the experimental and the fitted results, respectively.

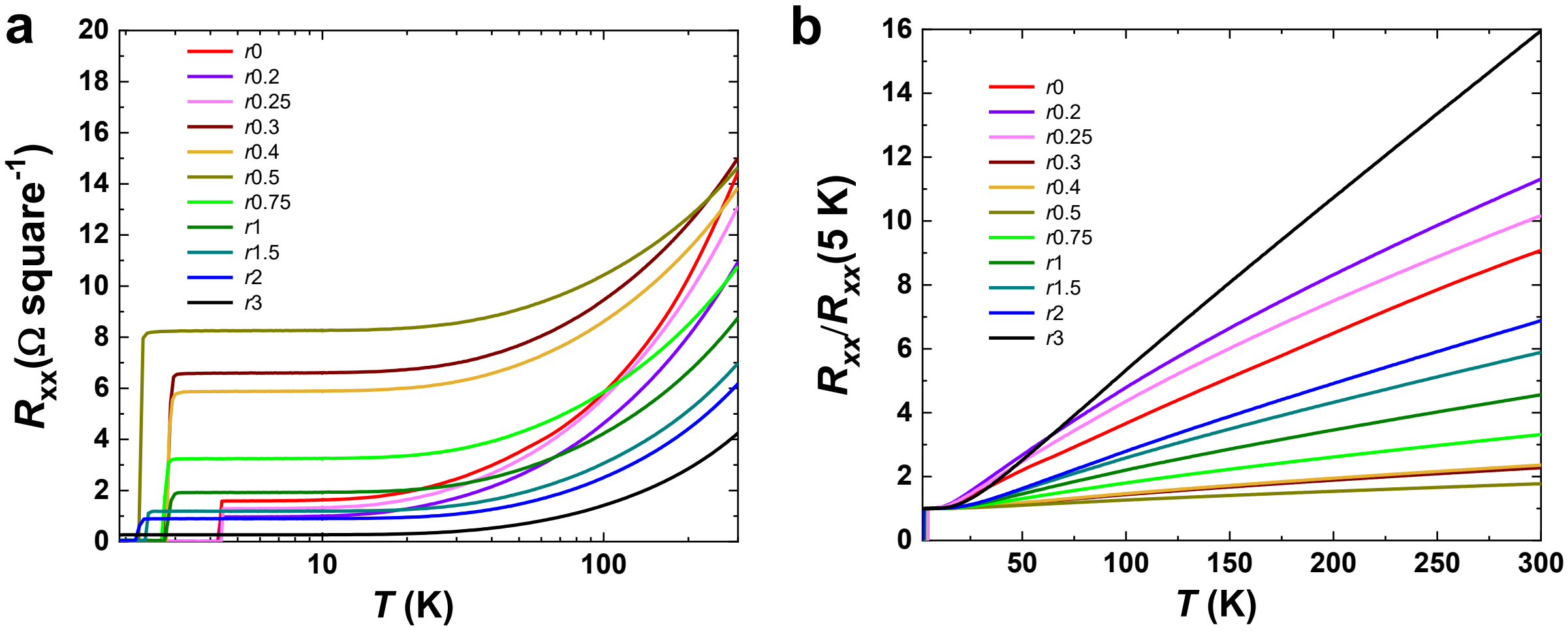

**Figure *S*5.** Temperature-dependent resistance of Pd-telluride films with varying *r* values. (a) $R_{xx}$ as a function of temperature for samples with various *r* values. (b) normalized resistance versus temperature, highlighting *RRR* across the samples.

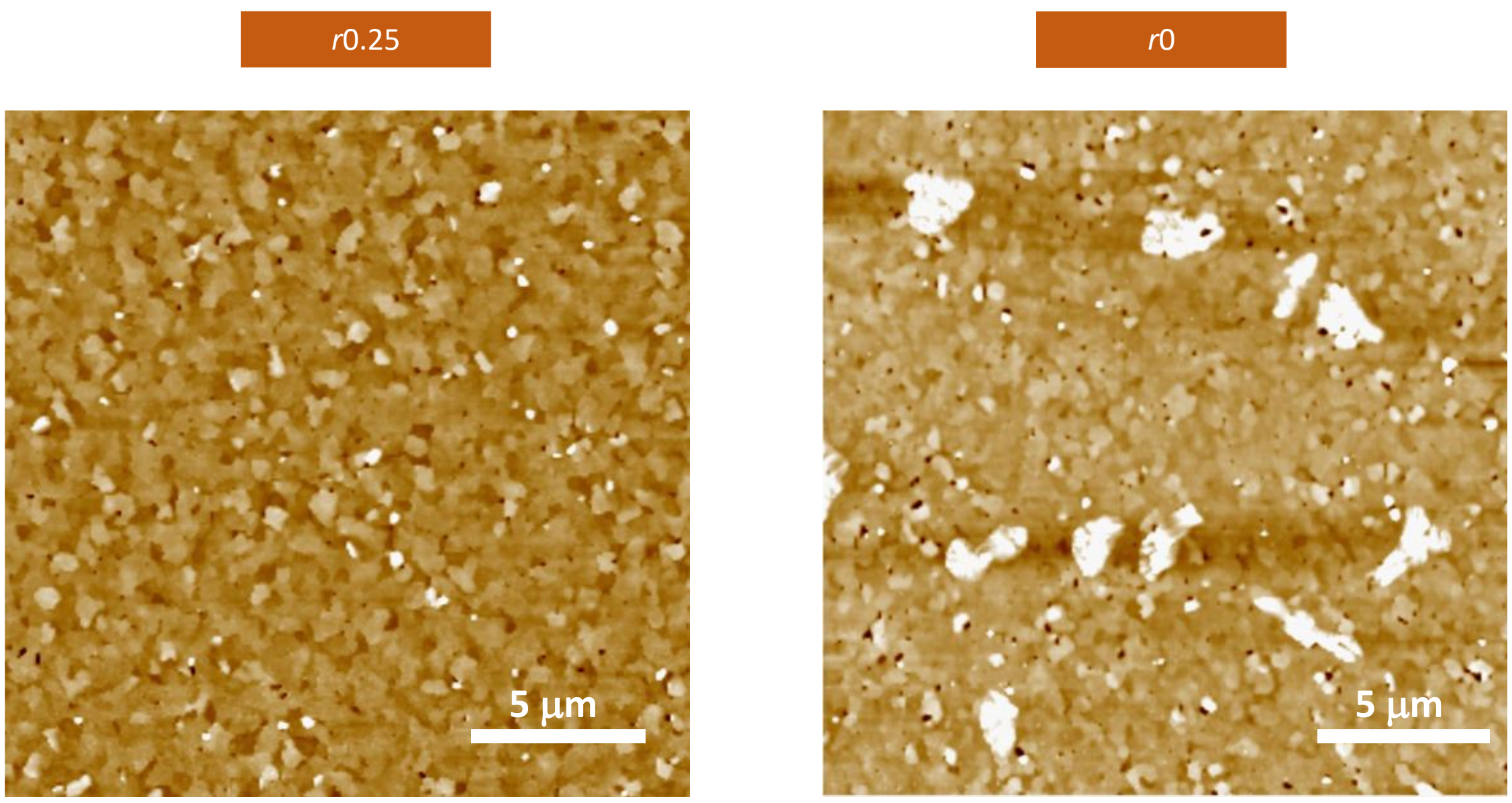

**Figure *S*6.** Atomic Force Microscopy (AFM) images of the surface for *r*0.25 and *r*0 samples with a scan size of 20 x 20 $\mu m^2$. The AFM image for the *r*0 sample displays clustered islands of excess Pd atoms.

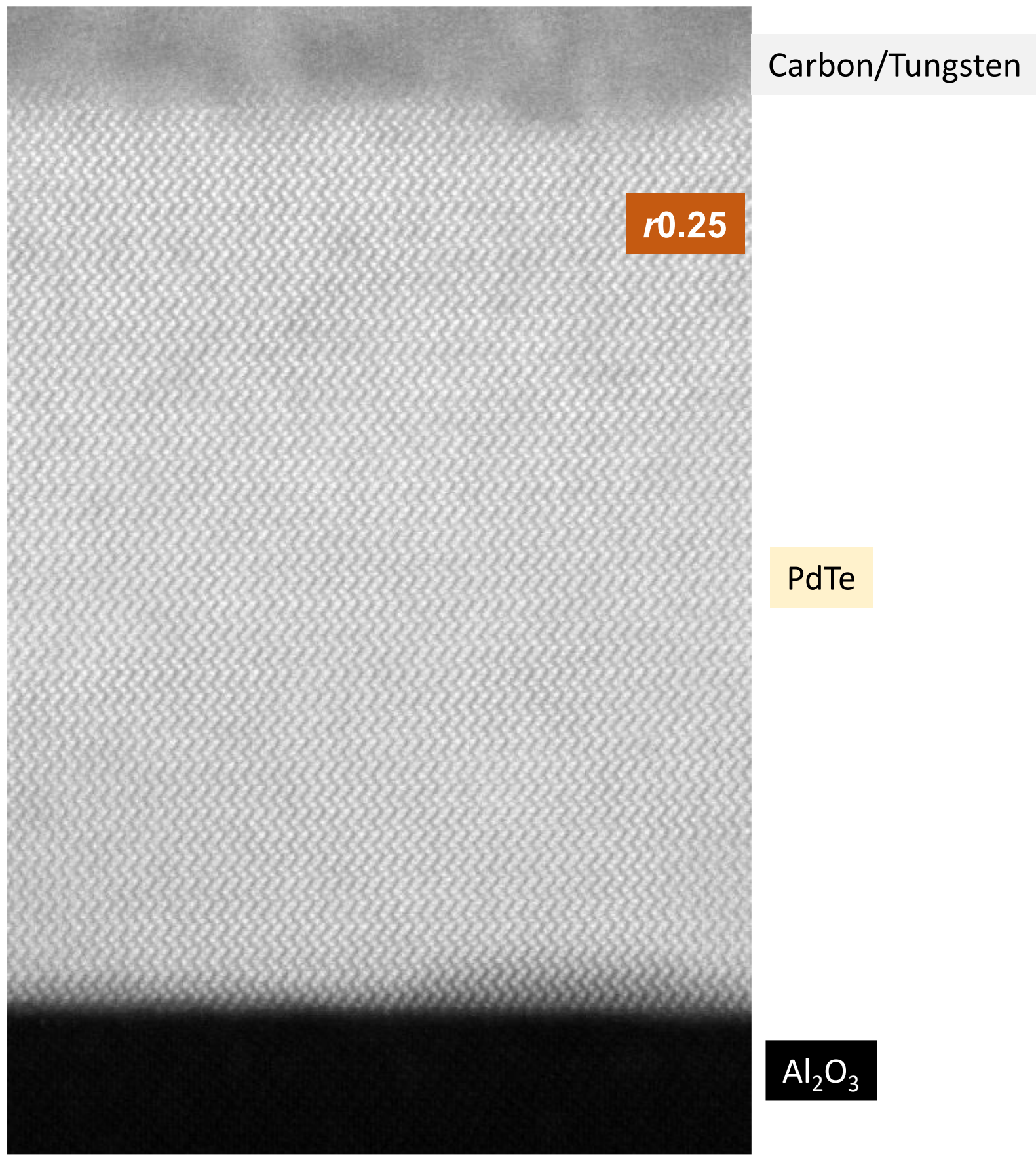

**Figure *S7*.** STEM image of Pd-telluride film with $r$ = 0.25 sample. The low-magnification cross-sectional image shows a zigzag pattern across the entire film, with a clear interface between PdTe and the $Al_2O_3$ substrate.